\documentclass[useAMS,usenatbib]{mn2e}
\pdfoutput=1

\usepackage{times} %use Adobe fonts - nicer for viewing PDF
\usepackage{aas_macros} % for journal abbreviations in Bibtex references from ADS 
\usepackage{amsmath}
\usepackage{amssymb} % for math symbols

\usepackage{graphicx}
\usepackage{epstopdf}
\usepackage{subfigure}

\newcommand{\bc}{\begin{center}}
\newcommand{\ec}{\end{center}}

\newcommand{\kms}{\mathrm{km \, s^{-1}}} 
\newcommand{\msol}{\mathrm{M_{\sun}}}
\newcommand{\vmax}{V_\mathrm{max}}

\title[The Milky Way system in $\Lambda$CDM cosmological simulations]
{The Milky Way system in $\Lambda$CDM cosmological simulations}

\author[Guo \textit{et al.}]{ \parbox{18cm}{Qi Guo$^{1,2}$, Andrew Cooper$^{2}$, Carlos Frenk$^{2}$,  John Helly$^{2}$, Wojciech Hellwing$^{2,3}$}\\
\\
$^{1}$ National Astronomical Observatories, Chinese Academy of Sciences, 20A Datun Road, Chaoyang, Beijing 10012, P.R. China\\
$^{2}$ Institute for Computational Cosmology, University of Durham, South Road, Durham DH1 3LE, UK\\
$^{3}$ Institute of Astronomy, University of Zielona G\'ora, ul. Lubuska 2, Zielona G\'ora, Poland}

\date{Accepted ... Received ... in original form ...}
\begin{document}
  
\maketitle

\begin{abstract} 
  We apply a semianalytic galaxy formation model to two high
  resolution cosmological N-body simulations to investigate analogues
  of the Milky Way system. We select these according to observed
  properties of the Milky Way rather than by halo mass as in most
  previous work. For disk-dominated central galaxies with stellar mass
  ($5$--$7)\times 10^{10}M_{\odot}$, the median host halo mass is
  $1.4\times10^{12}M_{\odot}$, with 1$\sigma$ dispersion in the range
  [0.86, 3.1]$\times10^{12}M_{\odot}$, consistent with dynamical
  measurements of the Milky Way halo mass. For any given halo mass,
  the probability of hosting a Milky Way system is low, with a maximum
  of $\sim$20\% in haloes of mass $\sim 10^{12}M_{\odot}$.  The model
  reproduces the $V$-band luminosity function and radial profile of
  the bright ($M_\mathrm{V} <  -9$) Milky Way satellites. Galaxy formation in
  low mass haloes is found to be highly stochastic, resulting in an
  extremely large scatter in the relation between M$_V$ (or stellar
  mass) for satellites and the depth of the subhalo potential well in
  which they live, as measured by the maximum of the rotation curve,
  $\vmax$. We conclude that the ``too big to fail'' problem is an
  artifact of selecting satellites in $N$-body simulations according to
  subhalo properties: in 10\% of cases we find that three or fewer of
  the brightest (or most massive) satellites have $\vmax> 30 \, \kms$. Our
  model predicts that around half of the dark matter subhaloes with
  $\vmax > 20 \,\kms$ host satellites fainter than $M_\mathrm{V} = -9$ and so may
  be missing from existing surveys.
\end{abstract}


\section{Introduction} 

The cold dark matter (CDM) model has been shown to be consistent with many
observations on cosmological scales, but uncertainties remain on small scales
where complex baryonic processes are involved. A substantial number of faint
satellite galaxies are known in the Milky Way system. Thanks to their relative
proximity, the distribution, chemistry and motions of their individual stars
can be measured precisely. This makes the Milky Way satellites an excellent
laboratory to test the CDM model and also to investigate the physics of galaxy
formation on small scales.

It has been known for some time that the count of substructures in
simulated Milky Way-mass dark matter haloes greatly exceeds that of the
known luminous Milky Way satellites
\citep[e.g.][]{Klypin1999,Moore1999}. In the context of CDM this is
readily explained if the efficiency with which baryons are converted
into stars drops quickly at low halo masses. This inefficiency is
expected on the basis of the well understood atomic physics governing
radiative cooling, the existence of an ionising cosmic UV background,
and comparison of the energy released by supernovae (which drive the
expulsion of baryons from dark matter haloes) to the depth of halo
potential wells \citep{Efstathiou_1992, Kauffmann_1993,
  Bullock_2000,Benson2002_sats,Somerville_2002}. CDM galaxy formation
models which incorporate even these most basic astrophysical effects
have been able to reproduce not only the abundance but also radial
distribution of satellites around the Milky Way
\citep[e.g.][]{Okamoto2010, Maccio_2010, Li10,
  Guo2011,Font2011,Parry2012}. A host of other
effects, such as cosmic ray pressure, may contribute to the regulation
of star formation in these most marginal systems,
\citep{Wadepuhl2011}.

Dwarf spheroidal galaxies (dSphs) make up most of the Milky Way (MW)
satellite population. Since they are dark matter-dominated galaxies,
they are ideal for testing the close connection between the properties
of dark matter haloes and the assembly of stellar mass predicted by CDM
models. Relatively direct comparison between models and observations
is now possible thanks to detailed kinematic analyses of the Milky Way
dSphs
\citep[e.g.][]{Penarrubia2008,Strigari2008,Lokas2009,Walker2009,Wolf2010,Strigari2010}.
\cite{Boylan2011,Boylan2012} compared observed dSph stellar kinematics
to predictions from the Aquarius Project, a set of six N-body
simulations of dark matter haloes of mass $\sim10^{12}\,\msol$
(consistent with constraints on the halo mass of the Milky Way; see
e.g. figure 1 of Wang et al. 2014). They concluded that the most
massive subhaloes in such simulations, which in typical galaxy
formation models would host the most luminous satellites, are always
too dense to be dynamically consistent with observations of any of the
known Milky Way companions. \cite{Boylan2011} dubbed this discrepancy
the ``too big to fail'' (TBTF) problem.

Several possible solutions to the problem have been advanced,
including alternative forms of dark matter \citep[e.g.][]{Lovell2012,
  Vogelsberger2013}, baryon-induced changes in dark halo density
profiles \citep{Cintio2011,Garrison2013,Arraki2013,Brooks2014}, and uncertainties
in the mass of the Milky Way halo \citep{Wang2012,Cautun2014}. All of
these studies use dark matter halo mass as a starting point to select
Milky Way analogues in simulations in order to analyze the internal
kinematics of their subhaloes.  However, galaxy formation involves many
complex processes and its outcome depends not only on the present-day
halo mass, but also on the formation history of the system. This is an
important consideration both for the selection of primary galaxies and
for the identification of relevant satellite haloes.  Indeed,
\cite{Sawala2014a} show that huge scatter exists in the relation
between the stellar mass of dwarf satellite galaxies and their
present-day subhalo mass. \cite{Sawala2014b} simulated analogues of
the Local Group including full baryonic physics and found that the
expulsion of baryons in winds from small dark matter haloes at early
times reduced the central density of the haloes in which satellites
form enough to solve the TBTF fail problem.

In this paper we implement the galaxy formation model of
\cite{Guo2013} on two N-body cosmological simulations and identify
Milky Way analogues using the properties of their central galaxies. We
compare the halo mass of these Milky Way analogues to the
observations, and examine the possibility of dark matter haloes of
given mass hosting the Milky Way analogues. We then further analyze
the abundance and kinematics of the subhaloes of satellite galaxies
selected by their luminosity or stellar mass. At the end we
investigate whether selecting MW analogues according to these 
observables could help solve the TBTF problem.
 
\section{Simulation and Semi-analytic Models}

This work makes use of two $\Lambda$CDM simulations: a high-resolution
zoom simulation, {\it Copernicus Complexio} (COCO; Hellwing et al. in
prep), and a cosmological volume simulation of lower resolution,
Millennium-II \citep[MS-II][]{Boylan2009}. The MS-II cube has
sidelength $100h^{-1}\,\mathrm{Mpc}$. The particle mass is
$m_p=6.9\times10^6 h^{-1} \msol$ and the force softening scale is
$\varepsilon= 1h^{-1} \mathrm{kpc}$. It assumes WMAP-1 cosmological
parameters, with a linear power spectrum normalization, $\sigma_8 =
0.9$, and matter density, $\Omega_m = 0.25$.

COCO simulates a high resolution comoving volume $V = 2.25\times 10^4
h^{-3}\;\mathrm{Mpc}^3$ in the centre of a lower resolution periodic
cosmological cube of sidelength $70.4 h^{-1} \mathrm{Mpc}$ using
$N_{\mathrm{p}}=2374^3$ DM particles. This results in a sample of
$\sim 50$ haloes of $\sim10^{12} \,\msol$ at $z=0$ with high mass
($m_p=1.135\times10^5 h^{-1} \msol$) and force ($\varepsilon=230
h^{-1} \mathrm{pc}$) resolution. Initial density perturbations were
generated with the novel {\it Panphasia} technique \citep{Jenkins2013}
which provides self-consistent and reproducible random phases for
zoom-in resimulations across an arbitrary range of
resolution\footnote{The specific COCO phase information is available
  on a request from the authors.}.

Unlike MS-II, COCO assumes WMAP-7 cosmological parameters
($\Omega_m=0.272$, $\Omega_{\Lambda}=0.728$, $\sigma_8=0.81$ and
$n_s=0.968$). The slightly different cosmogonies simulated by these
two calculations are reflected in the abundances and internal properties
of their dark matter haloes. For example, the abundance of MW-mass
haloes differs by a few per cent \citep{Boylan2011}. However, the
structural and kinematic properties of massive subhaloes in MW haloes
are not affected in any significant way \citep{Boylan2011, Wang2012,
  Garrison2013, Cautun2014}.

We used stored snapshots (160 for COCO and 68 for MS-II) for Friend-of-Friends
group finding \citep{Davis1985} with a linking length equal to $0.2$ times the
mean inter-particle separation. Subhaloes are identified by applying the SUBFIND
\citep{Springel2001} algorithm to each FOF group. We define the centre of the
FoF group as the potential minimum of the most massive self bound sub halo. The
viral mass, $M_{\mathrm{vir}}$, is defined as the mass enclosed by a virial
radius, $R_{\mathrm{vir}}$, which we approximate by the radius, $R_{200}$,
within which the mean density is 200 times the critical value for closure. The
maximum circular velocity of a halo is defined as $V_{\rm max} =
\mathrm{max}[\sqrt{GM(r)/r}]$ and is attained at radius $r_{\mathrm{max}}$.
For haloes close to the resolution limit,$\vmax$ can be systematically
underestimated. To correct for this effect, following \citet{Springel2008}, we
adjust measurements $\vmax^{\prime}$ from the simulation to $\vmax =
\vmax^{\prime}(1+(\varepsilon/r_{\mathrm{max}})^2)^{0.5}$.

To populate dark matter haloes in our two simulations with galaxies, we applied
the semi-analytic model recently developed by the Munich Group \citep{Guo2011,
Guo2013} to their merger trees. This model reproduces many statistical
properties of the observed galaxy population, including the abundance of bright satellites around the Milky Way
\citep{Guo2011}.  There are two processes crucial in understanding the
formation of low mass systems: UV reionization and supernova (SN) feedback.
\cite{Guo2011, Guo2013} adopted a fitting function originally proposed by
\cite{Gnedin2004} to describe the baryon fraction as a function of halo mass
and redshift. The fitting parameter and the characteristic halo mass beyond
which the baryon fraction is close the universal value but below which it
rapidly drops with decreasing halo mass are given by \cite{Okamoto2008}.

\cite{Guo2011, Guo2013} introduced a SN feedback model which depends strongly
on the maximum velocity of the subhalo, but the total amount of energy to
reheat and eject gas can never exceed the total energy released by SNII. For
most of the satellites around the Milky Way, the feedback energy saturates at
this value. \cite{Guo2011} also showed that SN feedback dominates the formation
of relative luminous satellites (M$_V <$ -11), while reionization becomes
important only for the very faint satellites ($\mathrm{M}_V >-11$).

In \cite{Guo2011}, satellite galaxies within the virial radius of their
host are subject to two environmental effects: stripping of their hot gas halo
by tides and ram pressure, and tidal disruption of their stellar component. The
orbit of the satellite's subhalo in the $N$-body simulation is used to estimate
these tidal and ram pressure forces. Once a satellite galaxy loses its parent
subhalo, it is considered to merge into the central galaxy unless the density
of the host halo at the pericenter of the satellite's orbit exceeds the average
baryonic mass density within the half mass radius of the satellite. In this latter case, cold gas from the disrupted satellite is added to the hot gas halo of the host, and stars from the satellite to a `stellar halo' component (which is not counted towards the stellar mass of the central galaxy).

 In the following sections we describe how we select MW analogues from the resulting mock catalogues and present our analysis of their satellite galaxy populations.

\section{Results}

The particle mass resolution of MS-II is sufficient to study the formation
history of haloes more massive than $\sim 10^{10}\,\msol$. Moreover, in our
semi-analytic model, the properties of central galaxies in haloes of mass
$10^{12}\,\msol$ are converged at the resolution of MS-II. However, the
resolution of MS-II is not sufficient to study the internal kinematics of
satellites with $\vmax \lesssim 30\;\kms$ \citep{Wang2012}. The resolution
limit of COCO corresponds to $\vmax \sim 10\;\kms$ (Hellwing et al. in prep),
corresponding to the least massive satellite galaxies known in the Milky Way
system. The smaller volume of COCO yields relatively few Milky Way analogues,
however. We therefore use the larger volume of MS-II to study the halo mass
distribution of Milky Way analogues selected according to their observable
properties, and the higher resolution of COCO to study the internal properties
of subhaloes in systems selected in this way.

\subsection{Halo mass of the Milky Way}

In previous works Milky Way analogues have been identified by
selecting isolated haloes with mass in a narrow range, typically around
$\sim 10^{12}M_{\odot}$. However, both theoretical work
\citep[e.g.][]{Guo2010, Moster2010, Behroozi2010, Cautun2014} and
observational data \citep[e.g.][]{Mandelbaum2006, Leauthaud2012}
suggest that the scatter in the relation between galaxy mass and host
halo mass is large. This raises the question of whether or not a
narrow halo mass range is sufficiently representative of the `Milky
Way' galaxy population.

In the following we use galaxy properties predicted by our semi-analytic
model to select Milky Way analogues, rather than dark matter halo mass. Our
selection is based the stellar mass and morphology of the central galaxy with
the following criteria:

\begin{equation} 5 \times 10^{10}M_{\odot} < M_* < 7\times 10^{10}
  M_{\odot} 
\end{equation} 
and 
\begin{equation} 0.05 <
  M_\mathrm{bulge}/{M_{*}} < 0.4, \end{equation}

where $M_{*}$ is the total stellar mass of the galaxy and
$M_{\mathrm{bulge}}$ is the stellar mass of the galactic bulge. We
have chosen this stellar mass range based on recent observational
constraints \citep[e.g.][]{McMillan2011}. The bulge selection is only
intended to restrict the sample to disk dominated galaxies. The actual
bulge mass fraction of the Milky Way is uncertain, as is the extent to
which the model `bulge' mass corresponds to the photometrically or
kinetically classified `bulges' of real disk-dominated galaxies. This
joint selection yields 566 Milky Way-like galaxies in MS-II and 13 in
COCO.

\begin{figure} 
\includegraphics[width=84mm, trim=0cm 1cm 6cm 8cm ]{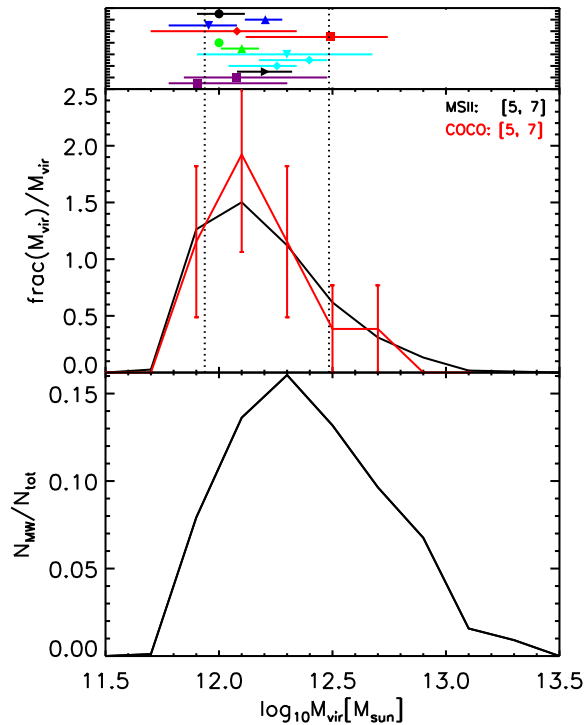} 
\caption{Upper panel: collection of measurements of $M_{vir}$ for
  the Milky Way taken from the literature (see text for
  details). Middle panel: the distribution of host halo mass for model
  galaxies with $5<M_{\star}<7 \times 10^{10}\msol$ and $0.05 <
  M_\mathrm{bulge}/{M_{*}} < 0.4$ selected in the MS-II (black curves)
  and COCO (red curves) simulations. Error bars show the Poisson
  errors for each mass bin. A KS test supports the hypothesis of a
  common parent distribution in the MS-II and COCO samples at a
  confidence level of $0.97$. Dotted lines mark the $16^{\mathrm{th}}$
  to $84^{\mathrm{th}}$ percentile range of the MS-II sample. Bottom
  panel: the probability for a halo of a given mass to host a galaxy
  that matches our Milky Way criteria in MS-II.}
\label{fig:MWmass} 
\end{figure}

In Fig.~\ref{fig:MWmass}, the middle panel shows the distribution of
$M_{\mathrm{vir}}$ for simulated Milky Way analogues selected according to
these criteria. The black curve represents the MS-II sample and the
red curve the COCO sample. Although the mass and spatial resolution of
the two simulations differ greatly, a K-S test supports the hypothesis
that these two samples are drawn from a common parent distribution at
a confidence level of 0.97 (roughly within Poisson errors, as shown).

The median halo mass is $1.4\times 10^{12}M_{\odot}$, with the
$16^{\mathrm{th}}$ to $84^{\mathrm{th}}$ percentile range (8.6$\times
10^{11}M_{\odot}$ to 3.1$\times10^{12}M_{\odot}$) indicated by black
dashed lines. A number of observational measurements of $M_{200}$ are
shown by symbols with error bars in the top panel of
Fig.~\ref{fig:MWmass}. These measurements have been made using the
kinematics of different halo tracers by \citet[][black
circle]{Xue2008}, \citet[][blue upward triangle]{Gnedin2010},
\citet[][blue downward triangle]{Watkins2010}, \citet[][cyan diamonds, with or without the inclusion of Leo I]{Sakamoto2003}
and \citet[][purple squares, assuming a truncated flat model or
NFW model]{Battaglia2005}; using the escape velocity of the Large
Magellanic Cloud or satellites \citep[][red
diamond]{Busha2011}; using masers by \citet[][green
circle]{Klypin2002}, incorporating photometric and kinematic data by
\cite[][green upward triangle]{McMillan2011}; using abundance
matching by \citet[][cyan downward triangle]{Guo2010}; using the
escape velocity of halo stars by \citet[][black rightfacing triangle]{Piffl2014}; and
using the timing argument of the Milky Way/Andromeda pair by
\citet[][red square]{Li2008}.  Almost all of the measurements
in the literature fall between the $16^{\mathrm{th}}$ and
$84^{\mathrm{th}}$ percentiles of the distribution predicted by our
model. The model predicts that haloes of even higher masses could also
host galaxies satisfying our Milky Way selection criteria: 31\% of the
host haloes in our samples are more massive than $2\times
10^{12}M_{\odot}$ and 17\% are more massive than $3\times
10^{12}M_{\odot}$.

We have tested how a systematic shift in the stellar mass of the Milky Way
affects the resulting halo mass distribution by assuming an alternative
stellar mass range, $4 \times 10^{10}M_{\odot} < M_* < 6\times
10^{10}M_{\odot}$. Relative to the peak of the black curve, we find the
distribution of $M_{\mathrm{vir}}$ shifts to lower mass by $\sim0.1$ dex. In
other words, the host halo mass range does not change much if we reduce the
median `Milky Way' stellar mass by 20\%.

The bottom panel of Fig.~\ref{fig:MWmass} shows the fraction of all
dark matter haloes of a given mass that host a central galaxy
satisfying our Milky Way selection criteria. We find that this
fraction reaches a maximum of around 16\% at $\sim 2\times
10^{12}M_{\odot}$. Most of the haloes this mass either host a galaxy
substantially more or less massive than the Milky Way, or else host a
bulge-dominated galaxy. The fraction of Milky Ways drops rapidly at
lower halo masses, and more slowly at higher masses. In general, even
in the most favoured range of $M_{200}$, the probability for a halo of
any given mass to host a MW analogue is remarkably low. Hydrodynamic
simulations of disk galaxies often preselect haloes from cosmological
volumes for `zoom' resimulations studies based on their mass. Our
results indicate that such a selection will always be
inefficient. Clearly, other properties of haloes and their individual
formation histories also need to be taken into account
\citep{Sawala2014a}.

\subsection{Satellite galaxies around the Milky Way}

In the last section we have shown that our model predicts a typical
halo mass for Milky Way analogues in the same range as recent
measurements of the Milky Way galaxy itself. In this section, we will
focus on the properties and spatial distribution of satellite galaxies
around the Milky Way. As mentioned in Sec.~2, the mass resolution of
the COCO simulation is 60 times higher than that of MS-II, and the
spatial resolution is higher by a factor of 4.  More than 90\% of
satellites in Mikly Way systems with $\vmax>15\,\kms$ still have their
own subhaloes in COCO, which allows us to follow their evolution in
detail (see Appendix). This level of resolution is especially
important for studying the internal dynamics and structure of
satellite galaxies. For example, their rotation curves can be traced
reliably and hence $\vmax$ can be used as a robust measure of the
depth of their potential. Therefore, in the following sections, we
only use the COCO simulation to study the properties of satellite systems.

\subsubsection{Abundance and profile}

\begin{figure}
\includegraphics[width=84mm, trim=2cm 2cm 3cm 8cm]{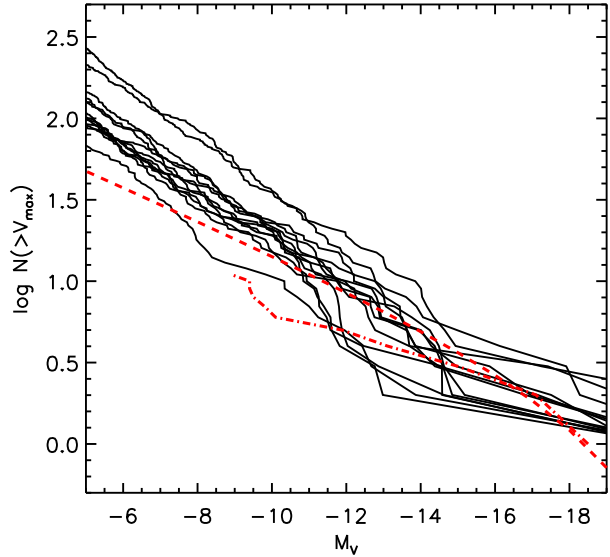}
\caption{Cumulative distribution of V-band magnitude for satellite
  galaxies in each simulated MW system (black curves).  The red
  dot-dashed line shows the equivalent distribution for known Milky
  Way satellites and the red dashed line the average distribution for
  Milky Way and M31 satellites corrected for completeness and sky
  coverage by \citet{Koposov2008}.}
\label{fig:LF}
\end{figure}

Fig.~\ref{fig:LF} shows cumulative counts of satellite galaxies as a
function of V-band magnitude around the 13 simulated MW analogues
selected from COCO (black curves). The red dot-dashed curve represents
the equivalent luminosity function of known satellites around the
Milky Way and the dashed curve shows the average luminosity function
of satellites in the MW and M31 found by \citet{Koposov2008}, who
attempted to correct for incompletness and partial sky coverage. For
this comparison, following Koposov et al., we define all galaxies
within 280~kpc of each MW analogue as its satellites. The model
predictions are broadly consistent with the data although the slope at
the faint end is steeper in the simulations than in the data. This
demonstrates convergence with the results of \citet{Guo2011}, who
showed that the bright end of the V-band satllite luminosity function
for the much larger number of MW analogues in MS-II is also in
reasonable agreement with the MW and M31 systems\footnote{The
  convergence of the MS-II MW satellite luminosity functions,
  demonstrated here by comparison to COCO, suggests that the
  semi-analytic treatment of orbits for galaxies whose dark matter
  haloes are stripped below the resolution of the $N$-body simulation
  works reasonably well.}.

\begin{figure}
\includegraphics[width=84mm, trim=2cm 2cm 3cm 6cm]{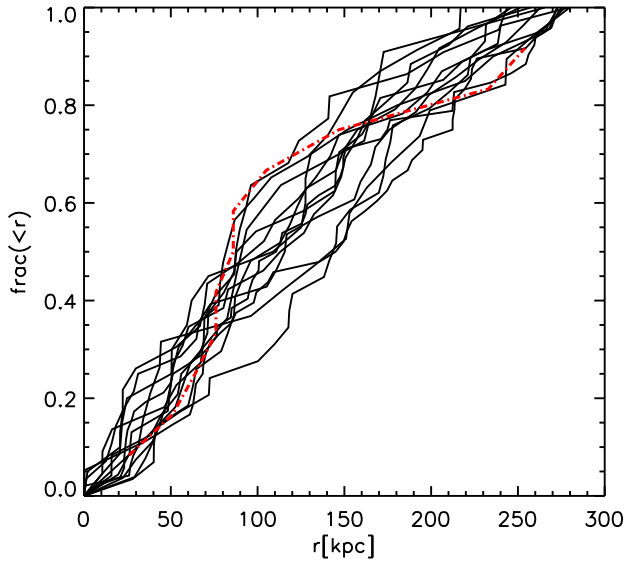}
\caption{The radial profile of the 12 brightest Milky Way satellite
  galaxies (red dashed curve, McConnachie et al. 2012) compared to the
  profiles of satellites with $M_V < -9$ in our COCO Milky Way
  analogue systems (black curves).}
\label{fig:profile}
\end{figure}

Fig.~\ref{fig:profile} compares the galactocentric radial distribution of the
most luminous satellites around our Milky Way analogues to those in the real
Milky Way system.  We restrict this comparison to the `classical' brightest 12
satellites, using the galactocentric distances given by \cite{McConnachie2012}. These
observed satellites galaxies are brighter than $M_{V} \approx -9$, so we select
simulated satellite galaxies with $M_\mathrm{V} < -9$ for this comparison. The
Milky Way observations lie within the envelope of the radial distributions from
our model.

\subsubsection{Distribution of $\vmax$}

In the following we focus on the kinematic properties of satellite
galaxies. Recently, detailed kinematic measurements have been carried
out for dSphs in the Local Group
\citep[e.g][]{Penarrubia2008,Strigari2008,Lokas2009,Walker2009,Wolf2010,Strigari2010,Strigari2014},
which makes it possible to perform a relatively direct comparison
between data and simulations. In particular, stellar kinematics are
often used to infer $\vmax$, which is then used to estimate
the gravitational potentials of satellite host dark matter haloes.

$\vmax$ is not a direct observable, of course, and must be inferred
from other properties, such as the mass within a given radius (usually
the mass within the half-mass radius, $M_{1/2}$), which in turn is
computed from a measured stellar velocity dispersion. To calibrate this
procedure, \cite{Boylan2012} computed $\vmax$ using subhaloes from the
Aquarius suite, a series of very high resolution resimulations of dark
matter haloes with $M_{200} \sim 10^{12}\,\msol$. Assuming that the
simulated subhaloes together constitute a representative sample of
Milky Way-like satellite galaxy hosts, they constructed a theoretical
distribution of $\vmax$ for each real MW satellite. This was done by
assigning a weight to the $\vmax$ of each simulated subhalo, according
to how closely its mass within a radius equal to the observed stellar
half-mass radius matched the value of $M_{1/2}$ inferred from
kinematic observations \citep[for details see][]{Boylan2012}.

\begin{figure}

 \includegraphics[width=84mm, trim=2cm 2cm 3cm 6cm]{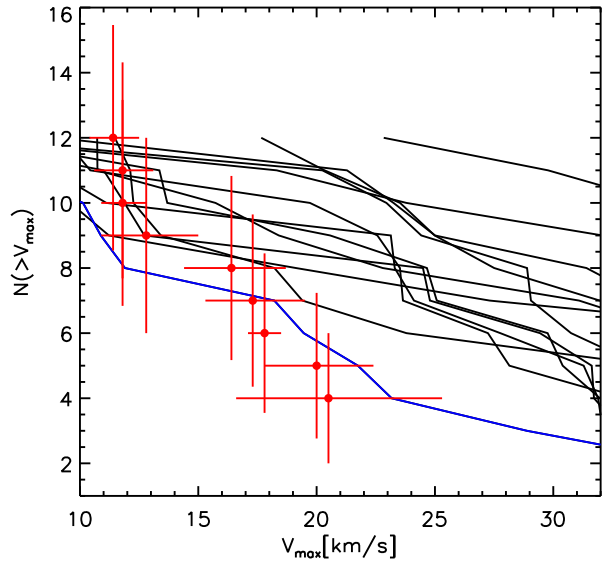}

 \caption{Cumulative distribution of maximum circular velocity for the
   12 most massive satellite galaxies in each simulated MW system
   ranked by stellar mass (blue and black curves). Red points show the
   values of $\vmax$ for MW satellites given by Boylan-Kochin et
   al. (2012). The vertical error bars on these points show the
   Poisson error.}
\label{fig:vmax}
\end{figure}

Fig.~\ref{fig:vmax} shows the cumulative distribution of $\vmax$ for
the 12 classical brightest satellites around the Milky taken from
Table~2 of \citet[][red symbols with error bars)]{Boylan2012}. For
comparison to this data, we rank the most massive 12 satellite galaxies of each model Milky Way
system according to their stellar mass and construct $\vmax$
distributions (black and blue lines).  The lowest value of $\vmax$
corresponding to resolved satellites in COCO is about $10\;\kms$ (see
Appendix).

Most of the simulated examples overpredict the abundance of satellites
with high $\vmax$ relative to the Milky Way. At
$\vmax\sim10$-$15\,\kms$, predicted subhalo abundances are consistent
with the MW system. At $\vmax\sim17\;\kms$, several model examples
still agree with the data within Poisson errors. One out of the 13
simulated Milky Way analogues has a $\vmax$ distribution consistent
with the data (within Poisson errors) over the whole magnitude range
considered here. We mark this system with a blue curve in
Fig.~\ref{fig:vmax}.

The host halo mass of the satellite system represented by the blue curve is
9.7$\times10^{11}M_{\odot}$, around the median of current estimates of the MW
host halo masses (Fig.~\ref{fig:MWmass}). Hence, this is not the least massive
halo in our sample. Of all our examples, this system has the lowest abundance
of satellites with high $\vmax$. This may seem somewhat at odds with the
conclusions of \citet{Wang2012} and \cite{Cautun2014}, who found that the
abundance of subhaloes at high $\vmax$ is proportional to the host halo mass.
Although such a relation holds on average, several factors introduce a large
amount of scatter into the correlation. In particular, the luminosity of dwarf
galaxies in our semi-analytic model depends strongly on their formation history
and not just on their present-day halo mass \citep[e.g.][]{Li10}. This broadens
the relation between luminosity and halo mass and hence that between luminosity
and $\vmax$. A similar scatter in the properties of low mass galaxies at a
fixed halo mass is also found in recent hydrodynamical simulations
\citep{Sawala2014a}.

\subsubsection{Too big to fail?}

\cite{Boylan2011} compared the abundance of dark matter subhaloes as a
function of $\vmax$ in the Aquarius simulations (assumed to be a
representative sample of MW host haloes) to the inferred $\vmax$
distribution of Milky Way satellites. They found that Aquarius
predicts significantly more subhaloes with high $\vmax$ compared to
what one could expect from all-sky extrapolation of the available
Milky Way data. Specifically, there are only 3 known satellites
galaxies with $\vmax$ above $30\,\kms$ (the Large and Small Magellanic
Clouds, and the tidally disrupted galaxy in Sagittarius), whereas
Aquarius predicts, on average, 8 subhaloes with $\vmax \gtrsim 30 \,
\kms$.  This is one of a number of ways to express the ``too big to
fail'' (TBTF) problem mentioned in the introduction.

\citet{Boylan2012} restated the TBTF problem by comparing the circular velocity
profiles of simulated subhaloes to the kinematically best-matching observed MW
dwarf spheroidal. Since the instantaneous value of $V_{\mathrm{max}}$ for a
given subhalo evolves over time (increasing as a subhalo grows, and decreasing
as it is tidally stripped), they compare with values
$V_{\mathrm{max}}(z_{\mathrm{ref}})$ at several different reference redshifts
($z_{\mathrm{ref}}=0, 10, z_{\mathrm{infall}}$, etc.). The idea is that the
present-day stellar mass in a subhalo should be tightly correlated with at
least one choice of $z_{\mathrm{ref}}$. As mentioned in the last section,
however, galaxy formation in low mass haloes can be highly stochastic. Here we
use the observables predicted by our model to test whether or not if suffers
from a TBTF problem.

%In \cite{Boylan2012} they use haloes of stellar mass [2.2, 1.0, 2.0, 2.2, 1.4, 1.3]$\times M_{\odot}$, which are within the one-$\sigma$ range of our predicted Milky Way mass.

\begin{figure}
\includegraphics[width=84mm, trim=2cm 2cm 3cm 6cm]{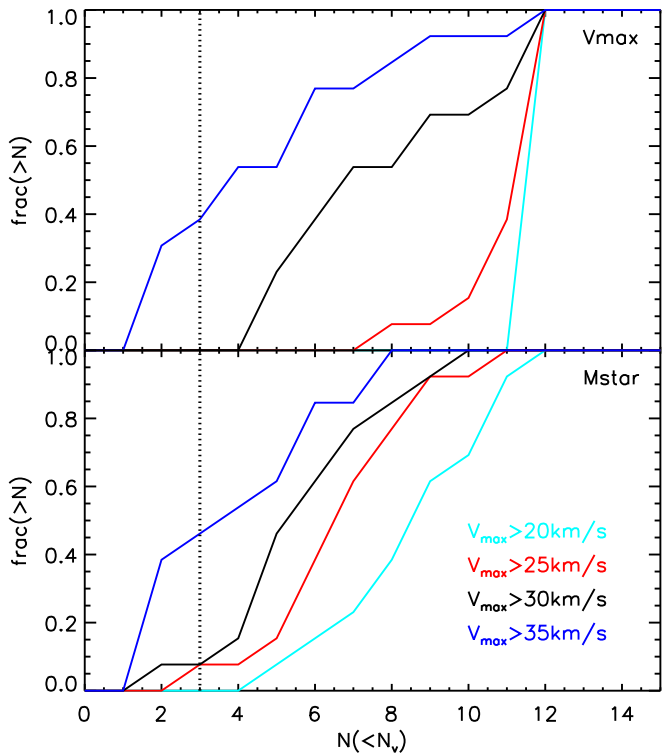}
\caption{Cumulative distribution of maximum circular velocity for the
  12 satellite galaxies with the highest $\vmax$ (upper panel) and
  stellar mass (lower panel) in each simulated MW system, averaged
  over all simulated MW systems. Different colours are for different
  $\vmax$ cuts as labelled in the right bottom corner in the bottom
  panel. Dotted vertical lines indicate the `critical number' of three satellites.}
\label{fig:tbtf}
\end{figure}

Fig.~\ref{fig:tbtf} shows the fraction of Milky Way analogues which
host $N$ or fewer satellites with $\vmax$ greater than a threshold
value (given in the legend). In the top panel, as \cite{Boylan2011,
  Boylan2012} did, we select the 12 subhaloes with the largest values of
$\vmax$ at $z=0$. We immediately see the fraction of Milky Way
analogues hosting no more than $N=3$ satellites galaxies with $\vmax
\gtrsim 30 \, \kms$ is zero, in line with the Aquarius simulations
discussed above. However, assuming approximate $\sqrt{N}$ shot noise
in the real Milky Way count, we can also note that the fraction of
hosts having no more than $N=5$ subhaloes with $\vmax \gtrsim 30 \,
\kms$ is 23\%.

The fraction of `MW-compatible' systems decreases at a given value of
$N$ if we lower the threshold $\vmax$ (red and cyan curves), and
increases if we raise the threshold (blue curve). A recent study
compared Local Group analogues in hydrodynamical and collisionless
simulations from the same initial conditions \citep{Sawala2014b} and
found that baryonic processes associated with feedback systematically
reduce $\vmax$ for a given subhalo by $\sim$ 10\% - 15\%. As our
simulation is collisionless, it is reasonable to take this effect into
account by assuming observed satellites with a measured $\vmax =
30\,\kms$ correspond to collisionless haloes with slightly higher
$\vmax$. This makes a significant difference -- for example, stating
the problem in terms of subhaloes with $\vmax \gtrsim 35 \,\kms$ yields a
fraction of `MW-compatible' hosts around 40\%.

The TBTF problem can be further alleviated by selecting simulated
satellites for comparison according to their observable properties,
rather than $\vmax$. In the bottom panel of Fig.~\ref{fig:tbtf} we
select the 12 satellites with the highest stellar mass and ask if this
sample also suggests a TBTF problem. The figure shows that, even
without considering the possibility that baryonic processes may reduce
the values of $\vmax$, 10\% of Milky Way analogues in our model have
three or fewer satellites with $\vmax < 30 \,\kms$, compatible with
the real Milky Way. If systems with $N=5$ or fewer satellites are
considered comparable, then even a low $\vmax$ threshold $20 \,\kms$
yields a compatible fraction of 10\% (this is a very conservative
threshold, since all 12 brightest satellites of the MW, except the
LMC, SMC and Sagittarius, have $V_{\mathrm{circ}} \lesssim 20\,\kms$
at their half light radius).

In summary, in addition to baryonic effects on satellite potentials
and the uncertainty in the appropriate statistical error bar on the
observed count of Milky Way satellites, the selection of simulated
satellite galaxies by observables increases the fraction of simulated
$\sim10^{12}\,\msol$ haloes that have abundances of `high $\vmax$'
satellites similar to that of the Milky Way system. Therefore,
although lowering the assumed host halo mass alleviates the TBTF
problem \citep{Vera2013, Wang2012, Cautun2014} this may not be
necessary if a self-consistent galaxy formation model, rather than a
$\vmax$ scaling relation, is used to identify the `brightest'
satellites. This approach may be crucial for formulating realistic
CDM-based theoretical predictions for MW-like satellite systems and
their properties.

\subsection{``Dark'' substructures}

The previous sections demonstrated that our model can reproduce the
abundance, radial profile and $\vmax$ distribution of the satellites
in the Milky Way, even though the abundance of subhaloes with high
$\vmax$ is larger than the count of bright satellites observed with
the same $\vmax$. In this section, we will show more quantitatively
the diverse properties of satellite galaxies at a given $\vmax$.

\subsubsection{$\vmax$ vs. luminosity}

\begin{figure}
\includegraphics[width=84mm, trim=2cm 2cm 3cm 6cm]{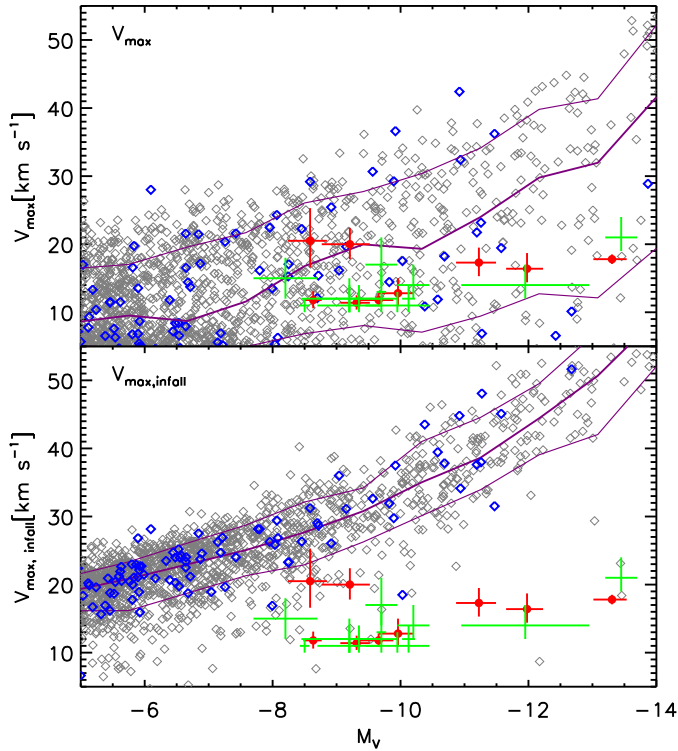}

\caption{Top panel: $\vmax$ vs. V-band luminosity for satellites of the Milky
Way. Grey diamonds are our model predictions. Blue diamonds highlight
satellites of the Milky Way analogue with a $\vmax$ distribution consistent
with the observed data over the whole magnitude range (blue curve in Fig. 4).
Thick and thin purple curves give the median and 16 to 84 percent range,
respectively.  Bottom panel: $V_\mathrm{max,infall}$ vs. V-band luminosity for
satellites of the Milky Way. The colour coding and line coding are the same as
in the upper panel. Red circles with error bars are the observed MW
satellites. Errors in magnitude are taken from Wolf et al.(2010), while those
in $\vmax$ are estimated by Boylan-Kolchin et al. (2012). Green symbols show
results for satellite galaxies in M31 as given by Tollerud et al. (2014). 
We do not include the SMC and LMC in this figure. $\vmax$ is $\sim 50 - 60 \,
\kms$ for the SMC \citep{Stanimirovic2004, Harris2006} and  $> 80 \, \kms$ for the LMC
(Olsen et al. 2011). For  Sagittarius, the appropriate value of $\vmax$ is hard
to determine, because the satellite appears to be strongly perturbed by its
interaction with the central potential of the Galaxy.} 

\label{fig:MstarVmax}
\end{figure}

The top panel of Fig.~\ref{fig:MstarVmax} shows the relation between
maximum circular velocity and V-band magnitude for satellites in our 13 Milky
Way analogues. The predicted median $\vmax$ is an increasing function of V-band
luminosity, but the scatter between $\vmax$ and magnitude, $M_{V}$, is very
large. Even haloes with $\vmax > 30$ or $20\,\kms$ can host galaxies fainter
than $M_{V} = -9$ (the approximate limits of completeness for current
all-sky surveys of the Milky Way) or $-5$, suggesting that a significant
number of $20-30\,\kms$ satellites are still to be discovered. Likewise, at a
given magnitude, $\vmax$ covers a wide range. For example, at $M_V \sim -12$,
$\vmax$ varies from $40\,\kms$ to $< 10\,\kms$.

This scatter between $\vmax$ and luminosity stems from the fact mentioned above
that the star formation rate history of a dwarf galaxy depends not only on its
final halo mass or $\vmax$, but also on its dark matter mass assembly history
\citet{Li10}. The bottom panel shows the relation between final stellar
mass and $\vmax$ measured just before the time of infall. This relation is much
tighter than the stellar mass vs. present-day  $\vmax$ relation shown in the
top panel.  This is because the subhaloes hosting satellite galaxies are
subject to tidal forces capable of stripping their mass. Galaxies embedded deep
inside the potential of the substructures are more condensed and thus more
resistant to such stripping, which can lower their central stellar
velocity dispersion without necessarily disturbing their surface brightness
profile \citep{Penarrubia2008, Cooper2010}.  Our model implies that in $1/10$
of MW analogues $\sim1-2$ relatively bright ($M_{V} \lesssim -13$) satellite
galaxies may have extremely low values of $\vmax$ (blue diamonds). In detail,
this number will depend sensitively on the depth at which stars are embedded in
their host potentials and on the details of tidal stripping.

Fig.~\ref{fig:MstarVmax} also shows the inferred relation between
$\vmax$ and luminosity for Milky Way and M31 satellites. $\vmax$ is
estimated in the same way in both cases \citep{Tollerud2014}. Most of
these data lie below the median prediction of the model (thick blue
curve), but well within the $16$--$84$ percentile
range. Interestingly, however, the most massive satellites in both
observational datasets lie furthest below the predicted median relation.

\subsubsection{Detection fractions}

In the last section we showed that the scatter between $\vmax$ and
luminosity is large, such that it is possible for relatively massive
subhaloes to host a galaxy with luminosity below the current all-sky
completeness limit. In this section we make an estimate of the
fraction of undetected subhaloes as a function of $\vmax$. For this
purpose, we consider $M_{V} = -9$ to be the current limit of
completeness and further assume that this is a hard cut. (In practice this
limit is also a function of surface brightness, and survey depth
varies across the sky; e.g. \citealt{Koposov2008}.)

\begin{figure}

\includegraphics[width=84mm, trim=2cm 2cm 3cm 6cm ]{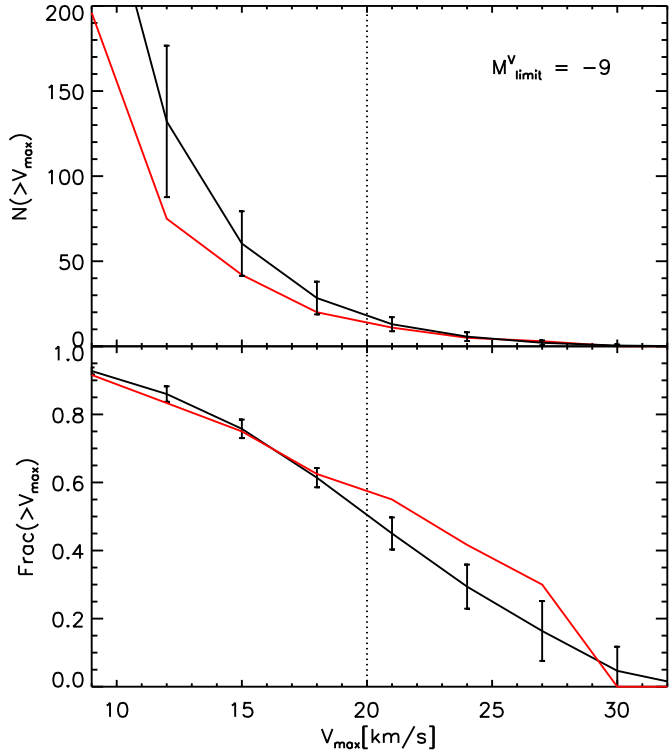}

\caption{Upper panel: number of subhaloes with galaxies fainter than
  $M_{V} = -9$ as a function of $\vmax$. The black curve shows the
  median value for all 13 Milky Way analogues and the error bars show
  the 16 to 84 percentile range. Red curves show the one system that
  is consistent with the observed Milky Way satellite $\vmax$ distribution, as
  shown in Fig.~\ref{fig:vmax}.  Bottom panel: fraction of the
  subhaloes hosting galaxies fainter than $M_{V} = -9$ as a function of
  $\vmax$. The colour coding is the same as in the upper panel.}
\label{fig:misssat}
\end{figure}

The top panel of Fig.~\ref{fig:misssat} shows the average number of
subhaloes hosting galaxies fainter than $M_{V} = -9$ as a function of
$\vmax$.  At $20\,\kms$ we find a median of 18 subhaloes below this
completeness limit. This corresponds to 45\% of all the dark matter
substructures of the same $\vmax$, as shown in the bottom panel. This
fraction increases rapidly with decreasing $\vmax$, such that around
90\% of subhaloes ($\sim250$) with $V_{max} > 10\,\kms$ are below the
limit. This conclusion holds for the single Milky Way analogue with a
$\vmax$ distribution consistent with Milky Way observations (red
curve in Fig.~\ref{fig:vmax}).

\section{Conclusions}

We have applied a semi-analytic galaxy formation model to cosmological N-body
simulations with sufficiently high resolution to study the properties of Milky
Way analogues and their satellite galaxies. Our two simulations give converged
predictions for the dark halo mass of the Milky Way, despite a large difference
in mass resolution (a factor of 60) and spatial resolution (a factor of 4).
Given its stellar mass, the most probable value for the halo mass of the Milky
Way according to our model is $1.4\times 10^{12}M_{\odot}$, with a
$1\sigma$ dispersion in the range $8.6\times10^{11}M_{\odot}$ to
$3.1\times10^{12}M_{\odot}$. This median value is consistent with measurements
of the Milky Way halo mass using different dynamical tracers \citep[see][for a
summary]{Wang2015}.

For any given halo mass, the probability of hosting a Milky Way
analogue is rather low, with a maximum probability of $\sim$ 20\% in
haloes of mass $\sim 10^{12} M_{\odot}$. This value decreases rapidly
with lower and higher host halo mass. This implies that if one wishes
to simulate the formation of a Milky Way like system, halo mass
selection alone is not sufficient: other factors such as environment
and assembly history are important. We intend to explore the influence
of such factors on the formation of Milky Way analogues in future
work.

We used the COCO simulation to study the properties of satellites in Milky Way
analogues. Our model is able to reproduce both the abundance of satellites as a
function of V-band luminosity and the radial profile of the bright galaxies
($M_\mathrm{V} < -9$). We have compared the $\vmax$ distribution of the
simulated satellites in COCO to inferences of $\vmax$ from observations. We
find one out of 13 $\vmax$ distributions for our Milky Way analogues overlaps
the observed distribution. However, we caution that the `observed' values of
$\vmax$ are themselves dependent on calibration against collisionless $N$-body
simulations.

% I don't think the following paragraph needs to appear in the conclusions. I moved the last sentence to the paragraph above.

% To study the dynamical structure of satellites requires simulations
% with resolution even higher than COCO. We thus use a characteristic
% quantity $\vmax$ to represent the dynamics of the satellites. This
% quantity cannot be observed directly but to be inferred taking
% advantage of some very high resolution simulations of individual
% haloes and assuming those substructures represent that of the Milky
% Way satellites. The $\vmax$ distribution of the predicted satellites
% in the COCO Milky Way analogue are compared to thus inferred
% $\vmax$. We find in 1 out of 13 Milky Way analogues the $\vmax$
% distribution overlaps with those observed.

With this caveat, a conservative upper limit on $\vmax$ for most known
satellite galaxies (excluding the LMC, SMC and Sagittarius) is
$\sim30\,\kms$. The severity of the ``too big to fail'' problem
depends on a number of systematic uncertainties, such as the error
assumed on the count of known satellites, the effect of baryonic
processes on $\vmax$ \citep[][and references therein]{Sawala2014b} and
the strength of the correlation between $\vmax$ and luminosity. Our
model shows that this last factor -- the selection of the comparison
sample according to luminosity in the presence of a large scatter
between luminosity and $\vmax$ -- is important for predicting the
number of MW satellites with high $\vmax$.

The large scatter between luminosity (stellar mass) and $\vmax$ is a
consequence of the complex formation histories of dwarf galaxies and
environmental effects (such as tidal and ram-pressure stripping) acting on dark
matter subhaloes. Subhaloes with high $\vmax$ can host very faint galaxies, and
bright galaxies can be hosted by haloes of low $\vmax$. A significant fraction
of subhaloes in our model host galaxies below the approximate all-sky
completeness limit for present surveys of $M_{\mathrm{V}} > -9$. Around half of our
subhaloes with $\vmax > 20 \kms$ host galaxies fainter than this limit.  The
continued discovery of fainter Milky Way companions by deeper surveys will be
important for constraining even the massive end of the subhalo mass function
and associated tests of the CDM model.

\section*{Acknowledgments}
We thank Jie Wang and Liang Gao for helpful discussion. QG acknowledges support from the NSFC grant (Nos. 11133003), the Strategic Priority Research Program ”The Emergence of Cosmological Structure” of the Chinese Academy of Sciences (No. XDB09000000), the ``Recruitment Program of Global Youth Experts'' of China, the NAOC grant (Y434011V01), MPG partner Group family, and a Royal Society Newton International Fellowship, as well as the hospitality of the Institute for Computational Cosmology at Durham University. APC is supported by a COFUND/Durham Junior Research Fellowship under EU grant [267209] and thanks Liang Gao for support in the early stages of this work under a CAS International Research Fellowship and NSFC grant [11350110323]. CSF acknowledges an ERC Advanced Investigator grant COSMIWAY [GA 267291]. This work was supported in part by the Science and Technology Facilities Council Consolidated Grant for Durham Astronomy [grant number ST/I00162X/1]. WAH is also supported by the Polish National Science Center [grant number DEC-2011/01/D/ST9/01960]. This work used the DiRAC Data Centric system at Durham University, operated by the Institute for Computational Cosmology on behalf of the STFC DiRAC HPC Facility (www.dirac.ac.uk). This equipment was funded by BIS National E-infrastructure capital grant ST/K00042X/1, STFC capital grant ST/H008519/1, and STFC DiRAC Operations grant ST/K003267/1 and Durham University. DiRAC is part of the National E-Infrastructure. The COCO simulations were run at the University of Warsaw HPC centre and we would like to thank Maciej Cytowski and Arkadiusz Niegowski for their help with these simulations. 
\bibliographystyle{mn2e}

\setlength{\bibhang}{2.0em} \setlength\labelwidth{0.0em}

\bibliography{draft} \

\begin{thebibliography}{}

\bibitem[\protect\citeauthoryear{{Arraki}, {Klypin}, {More} \&
  {Trujillo-Gomez}}{{Arraki} et~al.}{2013}]{Arraki2013}
{Arraki} K.~S.,  {Klypin} A.,  {More} S.,    {Trujillo-Gomez} S.,  2013, \mnras

\bibitem[\protect\citeauthoryear{{Battaglia}, {Helmi}, {Morrison}, {Harding},
  {Olszewski}, {Mateo}, {Freeman}, {Norris} \& {Shectman}}{{Battaglia}
  et~al.}{2005}]{Battaglia2005}
{Battaglia} G.,  {Helmi} A.,  {Morrison} H.,  {Harding} P.,  {Olszewski} E.~W.,
   {Mateo} M.,  {Freeman} K.~C.,  {Norris} J.,    {Shectman} S.~A.,  2005,
  \mnras, 364, 433

\bibitem[\protect\citeauthoryear{{Behroozi}, {Conroy} \& {Wechsler}}{{Behroozi}
  et~al.}{2010}]{Behroozi2010}
{Behroozi} P.~S.,  {Conroy} C.,    {Wechsler} R.~H.,  2010, \apj, 717, 379

\bibitem[\protect\citeauthoryear{Benson, Frenk, Lacey, Baugh \& Cole}{Benson
  et~al.}{2002}]{Benson2002_sats}
Benson A.~J.,  Frenk C.~S.,  Lacey C.~G.,  Baugh C.~M.,    Cole S.,  2002,
  MNRAS, 333, 177

\bibitem[\protect\citeauthoryear{{Boylan-Kolchin}, {Bullock} \&
  {Kaplinghat}}{{Boylan-Kolchin} et~al.}{2011}]{Boylan2011}
{Boylan-Kolchin} M.,  {Bullock} J.~S.,    {Kaplinghat} M.,  2011, \mnras, 415,
  L40

\bibitem[\protect\citeauthoryear{{Boylan-Kolchin}, {Bullock} \&
  {Kaplinghat}}{{Boylan-Kolchin} et~al.}{2012}]{Boylan2012}
{Boylan-Kolchin} M.,  {Bullock} J.~S.,    {Kaplinghat} M.,  2012, \mnras, 422,
  1203

\bibitem[\protect\citeauthoryear{{Boylan-Kolchin}, {Springel}, {White},
  {Jenkins} \& {Lemson}}{{Boylan-Kolchin} et~al.}{2009}]{Boylan2009}
{Boylan-Kolchin} M.,  {Springel} V.,  {White} S.~D.~M.,  {Jenkins} A.,
  {Lemson} G.,  2009, \mnras, 398, 1150

\bibitem[\protect\citeauthoryear{{Brooks} \& {Zolotov}}{{Brooks} \&
  {Zolotov}}{2014}]{Brooks2014}
{Brooks} A.~M.,  {Zolotov} A.,  2014, \apj, 786, 87

\bibitem[\protect\citeauthoryear{{Bullock}, {Kravtsov} \& {Weinberg}}{{Bullock}
  et~al.}{2000}]{Bullock_2000}
{Bullock} J.~S.,  {Kravtsov} A.~V.,    {Weinberg} D.~H.,  2000, \apj, 539, 517

\bibitem[\protect\citeauthoryear{{Busha}, {Marshall}, {Wechsler}, {Klypin} \&
  {Primack}}{{Busha} et~al.}{2011}]{Busha2011}
{Busha} M.~T.,  {Marshall} P.~J.,  {Wechsler} R.~H.,  {Klypin} A.,    {Primack}
  J.,  2011, \apj, 743, 40

\bibitem[\protect\citeauthoryear{{Cautun}, {Hellwing}, {van de Weygaert},
  {Frenk}, {Jones} \& {Sawala}}{{Cautun} et~al.}{2014}]{Cautun2014}
{Cautun} M.,  {Hellwing} W.~A.,  {van de Weygaert} R.,  {Frenk} C.~S.,  {Jones}
  B.~J.~T.,    {Sawala} T.,  2014, \mnras, 445, 1820

\bibitem[\protect\citeauthoryear{{Cooper}, {Cole}, {Frenk}, {White}, {Helly},
  {Benson}, {De Lucia}, {Helmi}, {Jenkins}, {Navarro}, {Springel} \&
  {Wang}}{{Cooper} et~al.}{2010}]{Cooper2010}
{Cooper} A.~P.,  {Cole} S.,  {Frenk} C.~S.,  {White} S.~D.~M.,  {Helly} J.,
  {Benson} A.~J.,  {De Lucia} G.,  {Helmi} A.,  {Jenkins} A.,  {Navarro} J.~F.,
   {Springel} V.,    {Wang} J.,  2010, \mnras, 406, 744

\bibitem[\protect\citeauthoryear{{Davis}, {Efstathiou}, {Frenk} \&
  {White}}{{Davis} et~al.}{1985}]{Davis1985}
{Davis} M.,  {Efstathiou} G.,  {Frenk} C.~S.,    {White} S.~D.~M.,  1985, \apj,
  292, 371

\bibitem[\protect\citeauthoryear{{di Cintio}, {Knebe}, {Libeskind}, {Yepes},
  {Gottl{\"o}ber} \& {Hoffman}}{{di Cintio} et~al.}{2011}]{Cintio2011}
{di Cintio} A.,  {Knebe} A.,  {Libeskind} N.~I.,  {Yepes} G.,  {Gottl{\"o}ber}
  S.,    {Hoffman} Y.,  2011, \mnras, 417, L74

\bibitem[\protect\citeauthoryear{{Efstathiou}}{{Efstathiou}}{1992}]{Efstathiou_1992}
{Efstathiou} G.,  1992, \mnras, 256, 43P

\bibitem[\protect\citeauthoryear{{Font}, {Benson}, {Bower}, {Frenk}, {Cooper},
  {De Lucia}, {Helly}, {Helmi}, {Li}, {McCarthy}, {Navarro}, {Springel},
  {Starkenburg}, {Wang} \& {White}}{{Font} et~al.}{2011}]{Font2011}
{Font} A.~S.,  {Benson} A.~J.,  {Bower} R.~G.,  {Frenk} C.~S.,  {Cooper} A.,
  {De Lucia} G.,  {Helly} J.~C.,  {Helmi} A.,  {Li} Y.-S.,  {McCarthy} I.~G.,
  {Navarro} J.~F.,  {Springel} V.,  {Starkenburg} E.,  {Wang} J.,    {White}
  S.~D.~M.,  2011, \mnras, 417, 1260

\bibitem[\protect\citeauthoryear{{Garrison-Kimmel}, {Rocha}, {Boylan-Kolchin},
  {Bullock} \& {Lally}}{{Garrison-Kimmel} et~al.}{2013}]{Garrison2013}
{Garrison-Kimmel} S.,  {Rocha} M.,  {Boylan-Kolchin} M.,  {Bullock} J.~S.,
  {Lally} J.,  2013, \mnras, 433, 3539

\bibitem[\protect\citeauthoryear{{Gnedin}, {Brown}, {Geller} \&
  {Kenyon}}{{Gnedin} et~al.}{2010}]{Gnedin2010}
{Gnedin} O.~Y.,  {Brown} W.~R.,  {Geller} M.~J.,    {Kenyon} S.~J.,  2010,
  \apjl, 720, L108

\bibitem[\protect\citeauthoryear{{Gnedin}, {Kravtsov}, {Klypin} \&
  {Nagai}}{{Gnedin} et~al.}{2004}]{Gnedin2004}
{Gnedin} O.~Y.,  {Kravtsov} A.~V.,  {Klypin} A.~A.,    {Nagai} D.,  2004, \apj,
  616, 16

\bibitem[\protect\citeauthoryear{{Guo}, {White}, {Angulo}, {Henriques},
  {Lemson}, {Boylan-Kolchin}, {Thomas} \& {Short}}{{Guo}
  et~al.}{2013}]{Guo2013}
{Guo} Q.,  {White} S.,  {Angulo} R.~E.,  {Henriques} B.,  {Lemson} G.,
  {Boylan-Kolchin} M.,  {Thomas} P.,    {Short} C.,  2013, \mnras, 428, 1351

\bibitem[\protect\citeauthoryear{{Guo}, {White}, {Boylan-Kolchin}, {De Lucia},
  {Kauffmann}, {Lemson}, {Li}, {Springel} \& {Weinmann}}{{Guo}
  et~al.}{2011}]{Guo2011}
{Guo} Q.,  {White} S.,  {Boylan-Kolchin} M.,  {De Lucia} G.,  {Kauffmann} G.,
  {Lemson} G.,  {Li} C.,  {Springel} V.,    {Weinmann} S.,  2011, \mnras, 413,
  101

\bibitem[\protect\citeauthoryear{{Guo}, {White}, {Li} \&
  {Boylan-Kolchin}}{{Guo} et~al.}{2010}]{Guo2010}
{Guo} Q.,  {White} S.,  {Li} C.,    {Boylan-Kolchin} M.,  2010, \mnras, 404,
  1111

\bibitem[\protect\citeauthoryear{{Harris} \& {Zaritsky}}{{Harris} \&
  {Zaritsky}}{2006}]{Harris2006}
{Harris} J.,  {Zaritsky} D.,  2006, \aj, 131, 2514

\bibitem[\protect\citeauthoryear{{Jenkins}}{{Jenkins}}{2013}]{Jenkins2013}
{Jenkins} A.,  2013, \mnras, 434, 2094

\bibitem[\protect\citeauthoryear{Kauffmann, White \& Guiderdoni}{Kauffmann
  et~al.}{1993}]{Kauffmann_1993}
Kauffmann G.,  White S. D.~M.,    Guiderdoni B.,  1993, MNRAS, 264, 201

\bibitem[\protect\citeauthoryear{{Klypin}, {Kravtsov}, {Valenzuela} \&
  {Prada}}{{Klypin} et~al.}{1999}]{Klypin1999}
{Klypin} A.,  {Kravtsov} A.~V.,  {Valenzuela} O.,    {Prada} F.,  1999, \apj,
  522, 82

\bibitem[\protect\citeauthoryear{{Klypin}, {Zhao} \& {Somerville}}{{Klypin}
  et~al.}{2002}]{Klypin2002}
{Klypin} A.,  {Zhao} H.,    {Somerville} R.~S.,  2002, \apj, 573, 597

\bibitem[\protect\citeauthoryear{{Koposov}, {Belokurov}, {Evans}, {Hewett},
  {Irwin}, {Gilmore}, {Zucker}, {Rix}, {Fellhauer}, {Bell} \&
  {Glushkova}}{{Koposov} et~al.}{2008}]{Koposov2008}
{Koposov} S.,  {Belokurov} V.,  {Evans} N.~W.,  {Hewett} P.~C.,  {Irwin} M.~J.,
   {Gilmore} G.,  {Zucker} D.~B.,  {Rix} H.-W.,  {Fellhauer} M.,  {Bell} E.~F.,
     {Glushkova} E.~V.,  2008, \apj, 686, 279

\bibitem[\protect\citeauthoryear{{Leauthaud}, {Tinker}, {Bundy}, {Behroozi},
  {Massey}, {Rhodes}, {George}, {Kneib} \& et al.}{{Leauthaud}
  et~al.}{2012}]{Leauthaud2012}
{Leauthaud} A.,  {Tinker} J.,  {Bundy} K.,  {Behroozi} P.~S.,  {Massey} R.,
  {Rhodes} J.,  {George} M.~R.,  {Kneib} J.-P.,    et al. 2012, \apj, 744, 159

\bibitem[\protect\citeauthoryear{{Li}, {De Lucia} \& {Helmi}}{{Li}
  et~al.}{2010}]{Li10}
{Li} Y.-S.,  {De Lucia} G.,    {Helmi} A.,  2010, \mnras, 401, 2036

\bibitem[\protect\citeauthoryear{{Li} \& {White}}{{Li} \&
  {White}}{2008}]{Li2008}
{Li} Y.-S.,  {White} S.~D.~M.,  2008, \mnras, 384, 1459

\bibitem[\protect\citeauthoryear{{{\L}okas}}{{{\L}okas}}{2009}]{Lokas2009}
{{\L}okas} E.~L.,  2009, \mnras, 394, L102

\bibitem[\protect\citeauthoryear{{Lovell}, {Eke}, {Frenk}, {Gao}, {Jenkins},
  {Theuns}, {Wang}, {White}, {Boyarsky} \& {Ruchayskiy}}{{Lovell}
  et~al.}{2012}]{Lovell2012}
{Lovell} M.~R.,  {Eke} V.,  {Frenk} C.~S.,  {Gao} L.,  {Jenkins} A.,  {Theuns}
  T.,  {Wang} J.,  {White} S.~D.~M.,  {Boyarsky} A.,    {Ruchayskiy} O.,  2012,
  \mnras, 420, 2318

\bibitem[\protect\citeauthoryear{{Macci{\`o}}, {Kang}, {Fontanot},
  {Somerville}, {Koposov} \& {Monaco}}{{Macci{\`o}} et~al.}{2010}]{Maccio_2010}
{Macci{\`o}} A.~V.,  {Kang} X.,  {Fontanot} F.,  {Somerville} R.~S.,  {Koposov}
  S.,    {Monaco} P.,  2010, \mnras, 402, 1995

\bibitem[\protect\citeauthoryear{{Mandelbaum}, {Seljak}, {Kauffmann}, {Hirata}
  \& {Brinkmann}}{{Mandelbaum} et~al.}{2006}]{Mandelbaum2006}
{Mandelbaum} R.,  {Seljak} U.,  {Kauffmann} G.,  {Hirata} C.~M.,    {Brinkmann}
  J.,  2006, \mnras, 368, 715

\bibitem[\protect\citeauthoryear{{McConnachie}}{{McConnachie}}{2012}]{McConnachie2012}
{McConnachie} A.~W.,  2012, \aj, 144, 4

\bibitem[\protect\citeauthoryear{{McMillan}}{{McMillan}}{2011}]{McMillan2011}
{McMillan} P.~J.,  2011, \mnras, 414, 2446

\bibitem[\protect\citeauthoryear{{Moore}, {Ghigna}, {Governato}, {Lake},
  {Quinn}, {Stadel} \& {Tozzi}}{{Moore} et~al.}{1999}]{Moore1999}
{Moore} B.,  {Ghigna} S.,  {Governato} F.,  {Lake} G.,  {Quinn} T.,  {Stadel}
  J.,    {Tozzi} P.,  1999, \apjl, 524, L19

\bibitem[\protect\citeauthoryear{{Moster}, {Somerville}, {Maulbetsch}, {van den
  Bosch}, {Macci{\`o}}, {Naab} \& {Oser}}{{Moster} et~al.}{2010}]{Moster2010}
{Moster} B.~P.,  {Somerville} R.~S.,  {Maulbetsch} C.,  {van den Bosch} F.~C.,
  {Macci{\`o}} A.~V.,  {Naab} T.,    {Oser} L.,  2010, \apj, 710, 903

\bibitem[\protect\citeauthoryear{{Okamoto}, {Frenk}, {Jenkins} \&
  {Theuns}}{{Okamoto} et~al.}{2010}]{Okamoto2010}
{Okamoto} T.,  {Frenk} C.~S.,  {Jenkins} A.,    {Theuns} T.,  2010, \mnras,
  406, 208

\bibitem[\protect\citeauthoryear{{Okamoto}, {Gao} \& {Theuns}}{{Okamoto}
  et~al.}{2008}]{Okamoto2008}
{Okamoto} T.,  {Gao} L.,    {Theuns} T.,  2008, \mnras, 390, 920

\bibitem[\protect\citeauthoryear{{Parry}, {Eke}, {Frenk} \& {Okamoto}}{{Parry}
  et~al.}{2012}]{Parry2012}
{Parry} O.~H.,  {Eke} V.~R.,  {Frenk} C.~S.,    {Okamoto} T.,  2012, \mnras,
  419, 3304

\bibitem[\protect\citeauthoryear{{Pe{\~n}arrubia}, {Navarro} \&
  {McConnachie}}{{Pe{\~n}arrubia} et~al.}{2008}]{Penarrubia2008}
{Pe{\~n}arrubia} J.,  {Navarro} J.~F.,    {McConnachie} A.~W.,  2008, \apj,
  673, 226

\bibitem[\protect\citeauthoryear{{Piffl}, {Scannapieco}, {Binney}, {Steinmetz},
  {Scholz}, {Williams}, {de Jong}, {Kordopatis} \& et al.}{{Piffl}
  et~al.}{2014}]{Piffl2014}
{Piffl} T.,  {Scannapieco} C.,  {Binney} J.,  {Steinmetz} M.,  {Scholz} R.-D.,
  {Williams} M.~E.~K.,  {de Jong} R.~S.,  {Kordopatis} G.,    et al. 2014,
  \aap, 562, A91

\bibitem[\protect\citeauthoryear{{Sakamoto}, {Chiba} \& {Beers}}{{Sakamoto}
  et~al.}{2003}]{Sakamoto2003}
{Sakamoto} T.,  {Chiba} M.,    {Beers} T.~C.,  2003, \aap, 397, 899

\bibitem[\protect\citeauthoryear{{Sawala}, {Frenk}, {Fattahi}, {Navarro},
  {Bower}, {Crain}, {Dalla Vecchia}, {Furlong}, {Helly}, {Jenkins}, {Oman},
  {Schaller}, {Schaye}, {Theuns}, {Trayford} \& {White}}{{Sawala}
  et~al.}{2014}]{Sawala2014b}
{Sawala} T.,  {Frenk} C.~S.,  {Fattahi} A.,  {Navarro} J.~F.,  {Bower} R.~G.,
  {Crain} R.~A.,  {Dalla Vecchia} C.,  {Furlong} M.,  {Helly} J.~C.,  {Jenkins}
  A.,  {Oman} K.~A.,  {Schaller} M.,  {Schaye} J.,  {Theuns} T.,  {Trayford}
  J.,    {White} S.~D.~M.,  2014, ArXiv e-prints

\bibitem[\protect\citeauthoryear{{Sawala}, {Frenk}, {Fattahi}, {Navarro},
  {Bower}, {Crain}, {Dalla Vecchia}, {Furlong}, {Jenkins}, {McCarthy}, {Qu},
  {Schaller}, {Schaye} \& {Theuns}}{{Sawala} et~al.}{2014}]{Sawala2014a}
{Sawala} T.,  {Frenk} C.~S.,  {Fattahi} A.,  {Navarro} J.~F.,  {Bower} R.~G.,
  {Crain} R.~A.,  {Dalla Vecchia} C.,  {Furlong} M.,  {Jenkins} A.,  {McCarthy}
  I.~G.,  {Qu} Y.,  {Schaller} M.,  {Schaye} J.,    {Theuns} T.,  2014, ArXiv
  e-prints

\bibitem[\protect\citeauthoryear{{Somerville}}{{Somerville}}{2002}]{Somerville_2002}
{Somerville} R.~S.,  2002, \apjl, 572, L23

\bibitem[\protect\citeauthoryear{{Springel}, {Wang}, {Vogelsberger}, {Ludlow},
  {Jenkins}, {Helmi}, {Navarro}, {Frenk} \& {White}}{{Springel}
  et~al.}{2008}]{Springel2008}
{Springel} V.,  {Wang} J.,  {Vogelsberger} M.,  {Ludlow} A.,  {Jenkins} A.,
  {Helmi} A.,  {Navarro} J.~F.,  {Frenk} C.~S.,    {White} S.~D.~M.,  2008,
  \mnras, 391, 1685

\bibitem[\protect\citeauthoryear{{Springel}, {Yoshida} \& {White}}{{Springel}
  et~al.}{2001}]{Springel2001}
{Springel} V.,  {Yoshida} N.,    {White} S.~D.~M.,  2001, \na, 6, 79

\bibitem[\protect\citeauthoryear{{Stanimirovi{\'c}}, {Staveley-Smith} \&
  {Jones}}{{Stanimirovi{\'c}} et~al.}{2004}]{Stanimirovic2004}
{Stanimirovi{\'c}} S.,  {Staveley-Smith} L.,    {Jones} P.~A.,  2004, \apj,
  604, 176

\bibitem[\protect\citeauthoryear{{Strigari}, {Bullock}, {Kaplinghat}, {Simon},
  {Geha}, {Willman} \& {Walker}}{{Strigari} et~al.}{2008}]{Strigari2008}
{Strigari} L.~E.,  {Bullock} J.~S.,  {Kaplinghat} M.,  {Simon} J.~D.,  {Geha}
  M.,  {Willman} B.,    {Walker} M.~G.,  2008, \nat, 454, 1096

\bibitem[\protect\citeauthoryear{{Strigari}, {Frenk} \& {White}}{{Strigari}
  et~al.}{2010}]{Strigari2010}
{Strigari} L.~E.,  {Frenk} C.~S.,    {White} S.~D.~M.,  2010, \mnras, 408, 2364

\bibitem[\protect\citeauthoryear{{Strigari}, {Frenk} \& {White}}{{Strigari}
  et~al.}{2014}]{Strigari2014}
{Strigari} L.~E.,  {Frenk} C.~S.,    {White} S.~D.~M.,  2014, ArXiv e-prints

\bibitem[\protect\citeauthoryear{{Tollerud}, {Boylan-Kolchin} \&
  {Bullock}}{{Tollerud} et~al.}{2014}]{Tollerud2014}
{Tollerud} E.~J.,  {Boylan-Kolchin} M.,    {Bullock} J.~S.,  2014, \mnras, 440,
  3511

\bibitem[\protect\citeauthoryear{{Vera-Ciro}, {Helmi}, {Starkenburg} \&
  {Breddels}}{{Vera-Ciro} et~al.}{2013}]{Vera2013}
{Vera-Ciro} C.~A.,  {Helmi} A.,  {Starkenburg} E.,    {Breddels} M.~A.,  2013,
  \mnras, 428, 1696

\bibitem[\protect\citeauthoryear{{Vogelsberger} \& {Zavala}}{{Vogelsberger} \&
  {Zavala}}{2013}]{Vogelsberger2013}
{Vogelsberger} M.,  {Zavala} J.,  2013, \mnras, 430, 1722

\bibitem[\protect\citeauthoryear{{Wadepuhl} \& {Springel}}{{Wadepuhl} \&
  {Springel}}{2011}]{Wadepuhl2011}
{Wadepuhl} M.,  {Springel} V.,  2011, \mnras, 410, 1975

\bibitem[\protect\citeauthoryear{{Walker}, {Mateo}, {Olszewski},
  {Pe{\~n}arrubia}, {Wyn Evans} \& {Gilmore}}{{Walker}
  et~al.}{2009}]{Walker2009}
{Walker} M.~G.,  {Mateo} M.,  {Olszewski} E.~W.,  {Pe{\~n}arrubia} J.,  {Wyn
  Evans} N.,    {Gilmore} G.,  2009, \apj, 704, 1274

\bibitem[\protect\citeauthoryear{{Wang}, {Frenk}, {Navarro}, {Gao} \&
  {Sawala}}{{Wang} et~al.}{2012}]{Wang2012}
{Wang} J.,  {Frenk} C.~S.,  {Navarro} J.~F.,  {Gao} L.,    {Sawala} T.,  2012,
  \mnras, 424, 2715

\bibitem[\protect\citeauthoryear{{Wang}, {Han}, {Cooper}, {Cole}, {Frenk},
  {Cai} \& {Lowing}}{{Wang} et~al.}{2015}]{Wang2015}
{Wang} W.,  {Han} J.,  {Cooper} A.,  {Cole} S.,  {Frenk} C.,  {Cai} Y.,
  {Lowing} B.,  2015, ArXiv e-prints

\bibitem[\protect\citeauthoryear{{Watkins}, {Evans} \& {An}}{{Watkins}
  et~al.}{2010}]{Watkins2010}
{Watkins} L.~L.,  {Evans} N.~W.,    {An} J.~H.,  2010, \mnras, 406, 264

\bibitem[\protect\citeauthoryear{{Wolf}, {Martinez}, {Bullock}, {Kaplinghat},
  {Geha}, {Mu{\~n}oz}, {Simon} \& {Avedo}}{{Wolf} et~al.}{2010}]{Wolf2010}
{Wolf} J.,  {Martinez} G.~D.,  {Bullock} J.~S.,  {Kaplinghat} M.,  {Geha} M.,
  {Mu{\~n}oz} R.~R.,  {Simon} J.~D.,    {Avedo} F.~F.,  2010, \mnras, 406, 1220

\bibitem[\protect\citeauthoryear{{Xue}, {Rix}, {Zhao}, {Re Fiorentin}, {Naab},
  {Steinmetz}, {van den Bosch}, {Beers}, {Lee}, {Bell}, {Rockosi}, {Yanny},
  {Newberg}, {Wilhelm}, {Kang}, {Smith} \& {Schneider}}{{Xue}
  et~al.}{2008}]{Xue2008}
{Xue} X.~X.,  {Rix} H.~W.,  {Zhao} G.,  {Re Fiorentin} P.,  {Naab} T.,
  {Steinmetz} M.,  {van den Bosch} F.~C.,  {Beers} T.~C.,  {Lee} Y.~S.,  {Bell}
  E.~F.,  {Rockosi} C.,  {Yanny} B.,  {Newberg} H.,  {Wilhelm} R.,  {Kang} X.,
  {Smith} M.~C.,    {Schneider} D.~P.,  2008, \apj, 684, 1143

\end{thebibliography}

\appendix
\section{Numerical considerations concerning the maximum circular velocity of satellites}
\label{sec:resolution}

When haloes fall into larger systems, they orbit around the potential
centres as substructures, losing some of their mass to tidal
stripping. Some substructures  survive for a long time, while
others may be disrupted shortly after infall. The finite
particle mass of N-body simulations sets a minimum mass below which
subhaloes cannot be resolved (in SUBFIND, subhaloes must contain at
least 20 particles). Since most subhaloes lose mass though tidal
stripping, many will eventually fall below this limit and be `lost'
from the simulation. The lower the resolution of the simulation, the
more rapidly subhaloes will be lost artificially as a result of this
numerical limit.  The identification of the surviving substructures is 
thus affected by numerical resolution. 

\begin{figure}

\includegraphics[width=84mm, trim=1cm 1cm 8cm 8cm ]{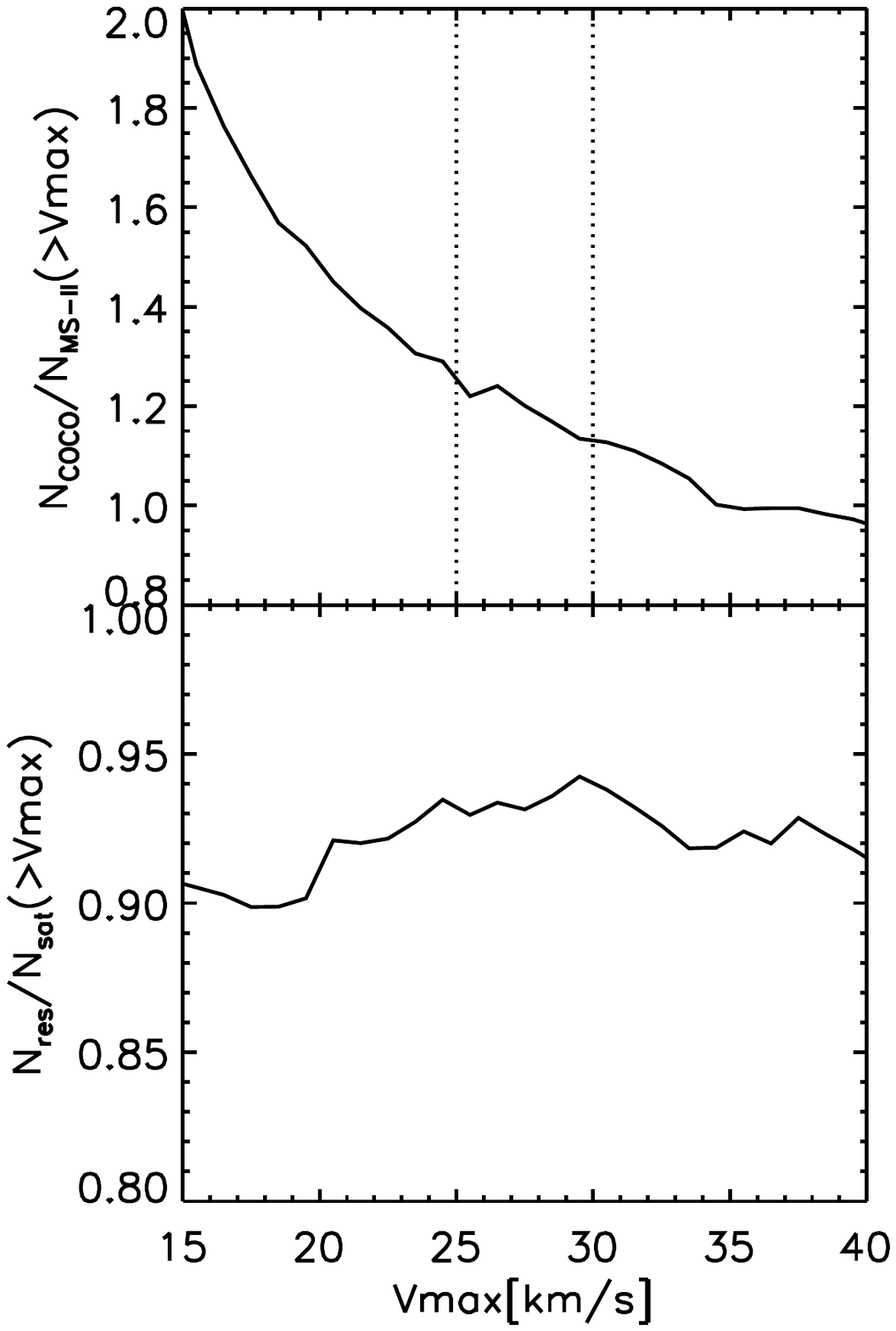}
\caption{Upper panel: the ratio of the cumulative number of subhaloes
  as a function of $\vmax$ in the COCO and MS-II simulations. Bottom:
  the ratio of the cumulative number of satellites with and without
  resolved subhaloes as a function of $\vmax$.}
\label{fig:VF}

\end{figure}

The MSII and the COCO simulations differ by a factor of 60 in particle
mass and by a factor of 4 in softening length. In Fig.~\ref{fig:VF} we
show the difference in the cumulative number of substructures as a
function of maximum circular velocity in the two simulations. The
difference is less than 20 percent at velocities, $\vmax > 25 \,
\kms$, but increases very rapidly to 100 percent at $15 \, \kms$. At
$20 \,\kms$, MS-II resolves only half of the subhaloes found in
COCO. We note that MS-II adopts a higher value of $\sigma_{8}$ which
results in more substructures. Hence, the curve in Fig.~\ref{fig:VF}
is a lower limit on the difference due to resolution. Studies of
satellite kinematics with MSII should be restricted to the regime
$\vmax > 30 \, \kms$, while most of the Milky Way satellites have
$\vmax < 20 \, \kms$, except for the LMC, SMC and Sagittarius
\citep[e.g][]{Boylan2012}. It is thus less reliable to use the MSII to
study the properties of satellite of the Milky Way without a very
careful treatment of these satellite galaxies whose subhaloes are lost due to
resolution.

The semianalytic galaxy formation model circumvents the resolution
limit by allowing satellite galaxies to survive as bound objects for
longer than their corresponding N-body subhaloes. We refer to these as
``orphan'' galaxies, because their parent subhalo has been lost. Their
continued survival is determined by analytic calculations of inspiral
towards the central galaxy under dynamical friction and the impact of
tidal stripping on their stellar profile. The value of $\vmax$ for
orphan galaxies is fixed to that of their parent subhalo at the time
it was last resolved. However, since these objects are suffering
severe tidal stripping by definition, this estimate of $\vmax$ is
unlikely to be very accurate.

The bottom panel of Fig.~\ref{fig:VF} shows the fraction of satellite galaxies
in our model whose subhaloes are resolved (i.e. are not orphans) as a function
$\vmax$. We find more than 90 percent of satellite galaxies in COCO are in
resolved subhaloes, and thus should have reliable $\vmax$ measurements. In the
main body of the paper we therefore use results from the COCO simulation to
study satellite galaxies.
  
\end{document}